\documentclass[aps,pra,nofootinbib,preprint,amsmath,amssymb,floatfix]{revtex4}
\usepackage{color}
\usepackage{graphicx}

\begin{document}

\newcommand{\tc}{\textcolor}
\newcommand{\g}{blue}
\newcommand{\ve}{\varepsilon}
\title{Classical and quantal aspects of Minkowski's four-momentum in analog gravity}         

\author{  Iver Brevik  }      

\affiliation{Department of Energy and Process Engineering, Norwegian University of Science and Technology, N-7491 Trondheim, Norway}
\date{\today}          

\begin{abstract}
The electrodynamic theory of continuous media is probably the most convenient platform when trying to construct  analog gravity theories. Quite naturally, this topic has gained considerable interest. One peculiar but not so very known feature in this context is the unconventional behavior of radiation energy and momentum in cases where superluminal fluid velocities are encountered, what, as   known, is  a major ingredient in analog gravity theories. These peculiar features are intimately connected with the spacelike character of Minkowski's four-momentum in electrodynamics. Here, we first consider an artificial model in which a Kerr-induced superluminal region is created in the right-hand region ($z>0$) in a left-moving, originally subluminal, fluid. We analyze the behavior of energy density, Poynting vector, and momentum density, and calculate the force on the artificial black hole horizon. Also, we  delve into quantal aspects, looking for eventual production of particles associated with the sudden creation of the horizon, finding,   however, that no particles are predicted to occur. The present paper continues a previous investigation by the  author on the same topic,  in Phys. Rev. A {\bf 100}, 032109 (2019). The subject as such  is closely related to the famous  Abraham-Minkowski problem.

\end{abstract}
\maketitle

\section{ Introduction}

Analog gravity is a subject that has, quite understandably, attracted considerable interest as it is aimed at demonstrating measurable effects mimicking at least to some extent  the characteristics of the real but so far non-accessible Hawking radiation in astrophysics \cite{hawking74}. The analog gravity picture has been introduced  in different areas of conventional physics, thus  both  in hydrodynamics where shallow water waves propagated in a strong counter current \cite{schutzhold02,rousseaux08}, in acoustics where a sonic-generated analog of a black hole horizon was analyzed \cite{unruh81}, and in electrodynamics with light propagating in moving dispersive media \cite{leonhardt00,drori19}. Readers interested in more extensive treatises on this topic, may consult the two volumes \cite{novello02,faccio13}, the article \cite{liao19} giving a lot of background material, and the useful recent survey over new analog gravity experiments \cite{jacquet20}.

We will in the present note deal with electromagnetic analog gravity only, and will focus on the following three points:

\noindent 1. The analog gravity theory is purely {\it classical.} It is right that in some situations, like the one analyzed by Drori {\i et al.} \cite{drori19}, one operates with emission frequencies in the emitter's rest inertial systems $\pm \omega$, thus giving the impression that one is mimicking the Hawking radiation from the horizon of a black hole. But the basic theory on the analog platform is nevertheless classical, and the experiment reported in Ref.~\cite{drori19} is an experiment entirely within classical electrodynamics.

\noindent 2. A key element in the theory is the {\it spacelike} property of Minkowski's four-momentum for the radiation field. It is precisely this property that makes it possible
to obtain a negative field energy in a class of inertial systems, and thus comes into play in analog gravity situations. From a basic viewpoint, the spacelike property results from the Minkowski momentum density  which in the rest system of matter is defined as
\begin{equation}
{\bf g}^{\rm M} = \bf D\times B. \label{0A}
\end{equation}
It is different from the Abraham momentum density which is defined as
\begin{equation}
  {\bf g}^{\rm A}   = \frac{1}{c^2}  \bf (E\times H). \label{0B}
\end{equation}
Thus  the relation between the momentum density and  Poynting's vector ${\bf S}={\bf E\times H}$ becomes
  ${\bf g}^{\rm M}=n^2{\bf S}/c^2$ with $n$ the refractive index.
 The usual  relationship ${\bf g}={\bf S}/c^2$,    conventionally   taken to express Planck's principle of inertia of energy, is thus broken.  This characteristic property of Minkowski's momentum (or more generally Minkowski's energy-momentum tensor) does not seem to be very well known, although the present author has pointed it out repeatedly, also in an analog gravity context \cite{brevik19}. (The basic presentation given in M{\o}ller's book \cite{moller72} is highly recommendable.)
 In all experiments on radiation pressure that we are aware of, the Minkowski energy-momentum tensor has turned out to be the simplest and most convenient alternative for  explaining  the measured effects \cite{brevik18}.

\noindent 3. As our third point, we will make a short excursion into quantum mechanics, not by following the stationary wave pattern in our toy model (it is specified below), but by assuming that the geometrical setup is imposed {\it suddenly}. Use of the Bogoliubov transformation allows on to estimate the produced number of toy-model scalar 'photons' in a simple way. This sort of calculation has been made use of repeatedly, for instance by  Parker \cite{parker87} when calculating the number of massless particles after the sudden formation of a cosmic string. Our result is, however, that the number of predicted particles is zero. Some simple physical considerations make it not so unreasonable that this should be so.

In the next section we continue the classical analysis of the subluminal-superluminal ideal fluid model introduced in Ref.~\cite{brevik19}, where an electromagnetic wave propagates against the current. In particular, we calculate the electromagnetic pressure on the artificial black horizon. To our knowledge, such a calculation has not been done before. In Sec.~III, we assume that the superluminal region $z>0$ is created instantaneously, applying the mentioned sudden approximation from quantum mechanics.  The physics is here not quite trivial, since a positive photon energy in the subluminar region acquires a negative value in the superluminal region. This is just the characteristic feature of the Minkowski four-momentum.

\section{A plane electromagnetic wave passing through the luminal barrier}

We will start from the same Gedanken experiment  as in Ref.~\cite{brevik19}, considering the   inertial frame $S$  in which a simple nondispersive  fluid is flowing with constant velocity $\bf v$ in the negative $z$ direction. We assume $|\bf v|$ large, but slightly less than the luminal limit $c/n$, where $n=\sqrt{\varepsilon_r \mu_r}>1$ with $\varepsilon_r$ and $\mu_r$ being  the relative  permittivity and the permeability of the medium when it is  at rest. Under these conditions it will thus be possible for a classical plane wave to propagate in the $z$ direction in the frame $S$,  from left to right.

To begin with, let us however keep the electromagnetic fields general, though classical. For readability, we   recall some elements of the basic theory:  in any inertial frame there are two field tensors, $F_{\mu \nu}$ and $H_{\mu \nu}$, related to the electric and magnetic fields via
$F_{ik}=B_l, \,  F_{4k}=(i/c)E_k, \, H_{ik}=H_l, \quad H_{4k}=ic D_k, \, (i,k,l$ cyclic). The Minkowski energy-momentum tensor can be expressed covariantly in the form
\begin{equation}
 S_{\mu\nu}= F_{\mu\alpha}H_{\nu\alpha}-\frac{1}{4}\delta_{\mu\nu} F_{\alpha\beta}H_{\alpha\beta}, \label{Minkowski}
\end{equation}
corresponding to the spatial components
\begin{equation}
S_{ik}=-E_iD_k-H_iB_k+\frac{1}{2}\delta_{ik}\bf{(E\cdot D+H\cdot B)}.
\end{equation}
This expression holds in all frames.
We make use of  SI units basically, but avoid writing out the vacuum constants $\varepsilon_0$ and $\mu_0$ explicitly to simplify the formulas (we will not deal with numerics in this paper). It may be noted that if one puts $c=1$, the formalism becomes identical to that following from the Heaviside-Lorentz convention.
It is  useful to note that the constitutive relations which in the rest system are ${\bf D}=\varepsilon {\bf E}$ and ${\bf B}=\mu {\bf H}$ can be written covariantly as
\begin{equation}
 \mu H_{\mu\nu}=F_{\mu\nu}-\kappa (F_{\mu\alpha}V_\nu-F_{\nu\alpha}V_\mu)V_\alpha, \label{constitutiverelation}
 \end{equation}
where $\kappa = (n^2-1)/c^2$. Here $V_\mu=({\bf V}, V_4)$ is the four-velocity of the medium, satisfying $V_\mu V_\mu=-c^2$. Often it is convenient to use the real quantity $V_0=-iV_4$ instead of $V_4$. (Note that the imaginary value of $V_4$ is merely a result of our metric conventions; we employ the Minkowski metric for which $x_4=ict$.)

It is worth noticing that Eq.~(\ref{constitutiverelation}) can be used to find the covariant governing equation for the four-potential $A_\mu$ in a straightforward  way. We restrict  ourselves to a radiation field without charges or currents. As is known, one half of Maxwell's equations can be written as $\partial_\nu H_{\mu\nu}=0.$ Thus, by inserting $F_{\mu\nu}= \partial_\mu A_\nu-\partial_\nu A_\mu$ into the right hand side of Eq.~(\ref{constitutiverelation}), the desired differential equation for $A_\mu$ follows.

We also need to observe the dispersion relation for a monochromatic wave with wave number $\bf k$. From Ref.~\cite{brevik19} we cite, for the quantity  $k_0=\omega/c$,
\begin{equation}
k_0=\frac{\kappa V_0({\bf k\cdot V}) \pm \sqrt{(1+\kappa V_0^2){\bf k}^2-\kappa ({\bf k\cdot V})^2}}{1+\kappa V_0^2}. \label{dispersionrelation}
\end{equation}
This expression is always real.

We have so far laid down the basic formalism. Return now to the Gedanken experiment above, where the large though subluminal  fluid velocity $\bf v$ is directed to the left. Let region I be the one for which  $z<0$, characterized by absence of the magnetic field. Let the incident plane wave be
\begin{equation}
E_x=E_Ie^{i(k_Iz-\omega t)}, \label{incident}
\end{equation}
where $k_I=k_z= \tilde{n}_I\omega/c$, the effective refractive index being
\begin{equation}
\tilde{n}_I=\frac{1+\kappa V_0^2}{n-\kappa V_0|\bf V|}. \label{initialrefractiveindex}
\end{equation}
It is worth noticing that Maxwell's equations can in the frame $S$ be written in the conventional form
\begin{equation}
{\bf \nabla \times E}_I=-\frac{\partial {\bf B}_I}{\partial t},
 \quad {\bf \nabla \times H}_I=\frac{\partial {\bf D}_I}{\partial t}, \label{maxwell1}
\end{equation}
\begin{equation}
{\bf \nabla \cdot D}_I=0, \quad {\bf \nabla \cdot B}_I=0, \label{maxwell2}
\end{equation}
where ${\bf D}_I=\tilde{\varepsilon}_I{\bf E}_I, \, $
${\bf B}_I=\tilde{\mu}_I{\bf H}_I$. With  $\beta=v_z/c$, the effective permittivity and permeability are
\begin{equation}
\tilde{\varepsilon}_I = \frac{\tilde{n}_I}{\mu}\frac{\tilde{n}_I+|\beta|}{1+\tilde{n}_I|\beta|},\label{permittivity}
\end{equation}
\begin{equation}
\tilde{\mu}_I=\mu \tilde{n}_I\frac{1+\tilde{n}_I|\beta|}{\tilde{n}_I+|\beta|},
\end{equation}
satisfying
\begin{equation}
\tilde{\varepsilon}_I\tilde{\mu}_I= \tilde{n}_I^2.
\end{equation}
The formal analogy with standard  electrodynamics in resting matter is striking.
We are not aware that this  correspondence could have been seen directly, without calculating.

The Minkowski energy density for the incident field is
\begin{equation}
W_I= \frac{E_I^2}{2c^2}\, \frac{{\tilde n}_I}{\mu}\, \frac{{\tilde{n}}_I+|\beta|}{1+{\tilde{n}}_I|\beta|}, \label{21}
\end{equation}
the Poynting vector is
\begin{equation}
S_I=W_I\frac{c}{\tilde{n}_I}= \frac{E_I^2}{2\mu c}\, \frac{{\tilde{n}}_I+|\beta|}{1+{\tilde{n}}_I|\beta|},
\end{equation}
and the Minkowski momentum density (superscript M from now on omitted) is
\begin{equation}
g_I=W_I \tilde{n}_I c =  \frac{E_I^2}{2c^3}\, \frac{{\tilde n}_I^2}{\mu}\, \frac{{\tilde{n}}_I+|\beta|}{1+{\tilde{n}}_I|\beta|}
\end{equation}
Assume now that a  strong electric field is applied in the region $z>0$ such that the refractive index $n_1$ in the fluid's rest system becomes larger than $n$ via the Kerr effect and satisfies the condition $n_1|\beta|>1$.  from left to right.  A  right-moving waves in the region $z>0$ will be dragged backwards, to the left.

First, we have to observe that there will occur a reflection of the incoming right-moving considered above, from the surface $z=0$. Maxwell's equations (\ref{maxwell1}) and (\ref{maxwell2}) are still valid for the reflected wave with components $E_R,B_R$, etc., but now with the effective refractive index
\begin{equation}
\tilde{n}_R=\frac{1+\kappa V_0^2}{n+\kappa V_0|{\bf V}|}. \label{23}
\end{equation}
The transmitted wave will be designated by the analogous subscript $T$. The boundary conditions for the field at the interface are $E_I-E_R=E_T, \, H_I+H_R=H_T$. From these, we obtain after some calculation the following expression for the relative transmitted field \cite{brevik19}
\begin{equation}
\frac{E_T}{E_I}= \frac{\mu_1}{\mu}\, \frac{(\tilde{n}_I+\tilde{n}_R)(1+\beta^2)+2(1+\tilde{n}_I\tilde{n}_R)|\beta|} {  (\tilde{n}_T+\tilde{n}_R)(1+\beta^2)+2(1+\tilde{n}_T\tilde{n}_R)|\beta|} \, \frac{1+{\tilde n}_T|\beta|}{1+\tilde{n}_I|\beta|}. \label{forhold}
 \end{equation}
The transmitted wave, still polarized in the $x~$direction, is in analogy to Eq.~(\ref{incident})
\begin{equation}
E_x=E_Te^{i(k_Tz-\omega t)},
\end{equation}
where the wave number is $k_T = {\tilde n}_T\omega/c$, the effective refractive index in this region being
\begin{equation}
\tilde{n}_T= \frac{1+\kappa_1V_0^2}{n_1-\kappa_1V_0|{\bf V}|} \label{transmitted}
\end{equation}
with $\kappa_1=(n_1^2-1)/c^2$. The  magnetic field polarized along the $y$ axis is
\begin{equation}
H_T= \frac{E_T}{\mu_1c}\, \frac{\tilde{n}_T+|\beta|}{1+\tilde{n}_T|\beta|}. \label{transmittedmagneticfield}
\end{equation}
As the interface is at rest in the inertial system $S$, the electromagnetic force does no work in passing the boundary region around $z=0$, and the angular frequency $\omega$ is the same  in both regions.

This point is actually somewhat nontrivial.  For comparison, we may first consider the passage of light through a dielectric fluid surface in classical optics, assuming   the surface to be  at rest. In the boundary region there is a volume force density $-\frac{1}{2}E^2{\bf \nabla}n^2$, always acting from the optical thicker to the optical thinner region.  This force is unable to do   work under the resting conditions, and so the frequency is left unchanged upon passage through the boundary.  By integrating the mentioned force density across the surface, one obtains the same surface pressure  as from taking the difference between the normal electromagnetic stress components on the two sides. In our case the situation is different as the fluid is moving through the boundary. There is still a nonvanishing force density in the boundary layer, being essentially as that given above,  but it is still unable to do work as boundary layer is very thin. In conclusion, we can also in the present case assume that the light frequency is the same on the two sides.

Expressed in terms of $E_T$, the energy density in the transmitted region is
\begin{equation}
W_T=\frac{E_T^2 \tilde{n}_T}{\mu_1c^2}\,\frac{\tilde{n}_T+|\beta|}{1+\tilde{n}_T|\beta |}, \label{energy}
\end{equation}
the Poynting vector is
\begin{equation}
S_T= \frac{E_T^2 }{\mu_1c}\,\frac{\tilde{n}_T+|\beta|}{1+\tilde{n}_T|\beta |}, \label{poynting}
\end{equation}
and the momentum density
\begin{equation}
g_T= \frac{E_T^2 \tilde{n}_T^2}{\mu_1c^3}\,\frac{\tilde{n}_T+|\beta|}{1+\tilde{n}_T|\beta |}. \label{momentum}
\end{equation}
At this place the spacelike property of Minkowski's four-momentum turns up. Let us for simplicity consider the case where the fluid velocity $|v_z|$ lies close to the luminal limit $|\beta|=1/n$. This limit is also the case of main interest. We can then effectively replace $V_0$ with $n/\sqrt \kappa$, and $|\bf V|$ with $1/\sqrt \kappa$. It is then apparent that the refractive index (\ref{transmitted}) diverges. To avoid the formal divergence, let us introduce the small but positive difference $\Delta n$ between the  refractive indices $n_1$ and $n$ and write
\begin{equation}
\tilde{n}_T=-\frac{\kappa c^2}{\Delta n},  \quad \Delta n =n_1-n.
\end{equation}
It is worth noticing that this expression is clearly defined, once $\Delta n$ is given, even in the exact luminal limit.
It thus follows from Eq.~(\ref{energy}) that $W_T<0$, and large in magnitude. The negativity demonstrates the spacelike character  of the four-momentum. On the other hand, both the energy flux density  and the momentum stay positive,
\begin{equation}
S_T=\frac{E_T^2n}{\mu_1 c},
\end{equation}
\begin{equation}
g_T= \frac{E_T^2n \tilde{n}_T^2}{\mu_1 c^3},
\end{equation}
the magnitude of $g_T$ being large. We now approximate $\mu_1$ with $\mu$, and omit the $\Delta n$ corrections in Eq.~(\ref{forhold}) to get
\begin{equation}
E_T=E_I.
\end{equation}
The transmitted magnetic field is according to Eq.~(\ref{transmittedmagneticfield}) equal to $H_T=E_In/(\mu c)$, and so the energy flux density remains unchanged upon  passage through $z=0$,
\begin{equation}
S_T=S_I  = \frac{E_T^2n}{\mu c}.
\end{equation}
This is a nontrivial point. It implies that in the luminal limit $|\beta|=1/n$ the reflected wave is zero.

Let us finally consider the surface force density (pressure) on the black hole horizon. This is of obvious physical interest, but has to our knowledge not been considered before within classical electrodynamics. To comply with more conventional notation, we introduce the Maxwell stress tensor as $T_{ik}
=-S_{ik}$. Then, the surface pressure  called $\sigma_z$ is written as
\begin{equation}
\sigma_z= T_{zz}(0+) -T_{zz}(0-).
\end{equation}
On the outer side $z=0+$,  $T_{zz}(0+)=-W_T$, which means in view of   Eq.~(\ref{energy}) that
\begin{equation}
T_{zz}(0+)  = \frac{E_T^2\kappa n}{2\mu}\, \frac{1}{\Delta n}. \label{x}
\end{equation}
There is thus a strong, outward-directed, force acting on this side. The positivity of the force is a consequence of  the negativity of the transmitted energy. To re-emphasize, the result (\ref{x}) is clearly defined, even in the  luminal limit $|\beta|= 1/n$.

On the inner side $z=0-$, we have analogously $T_{zz}(0-)=-W_I$, as there is no contribution from the reflected  wave in the luminal limit.   Here  we encounter  the  problem  that the  effective refractive index $\tilde{n}_I$, defined by Eq.~(\ref{initialrefractiveindex}), diverges in  this limit. It will be instructive to work out the analytical approximation that applies   when $|\beta|$ approaches $1/n$ from below. Introducing the nondimensional velocity difference $\Delta \beta$ via
\begin{equation}
|\beta|=\frac{1}{n}-\Delta \beta, \quad \Delta \beta >0,
\end{equation}
we obtain by expanding to the first order in $\Delta \beta$,
\begin{equation}
\tilde{n}_I=\frac{\kappa c^2}{n^2}\, \frac{1}{\Delta \beta}. \label{transmittedrefractiveindex}
\end{equation}
This  quantity is  large and positive. Using Eq.~(\ref{21}) we then obtain
\begin{equation}
T_{zz}(0-)= -\frac{E_I^2\kappa}{2\mu n}\, \frac{1}{\Delta \beta}.
\end{equation}
The inner pressure is $- T_{zz}(0-)$, and thus positive. The total pressure on the surface $z=0$ finally takes the form
\begin{equation}
\sigma_z= \frac{E_I^2\kappa}{2\mu}\left( \frac{n}{\Delta n} +\frac{1}{n\Delta \beta } \right).
\end{equation}
Both the outer and the inner pressures are acting in the same direction, what is unusual in classical electrodynamics. The outward directed force is strong, and is very sensitive with respect to the  increase $\Delta n$ of the refractive index and to  the deviation $\Delta \beta$ in the fluid velocity from the luminal limit.

\section{Quantal aspects: Is there a scalar particle production?}

So far, our considerations have been entirely classical. As mentioned above it  is however possible to associate our  model with quantum mechanics in a straightforward way, by  assuming  that the Kerr-generated superluminal region $z>0$ is created {\it suddenly}, and thereafter held constant. Use of the sudden approximation in quantum mechanics then enables us to estimate if  particles are produced. The method is well known, for instance from cosmology, in connection with the sudden creation of cosmic strings.  We recall the work of   Parker \cite{parker87}, and there are also several other references.

   Thus assume that the Kerr-generated increase $\Delta n$ of the refractive index in the region $z>0$ is produced at the instant $t=0$. For $t<0$ there is only an undisturbed fluid  moving subluminally  to the left, and the region $z>0$ is identical to the incident region called {\it I} in the previous section. Of physical interest are  the right-moving modes when $z>0$; the left-moving modes in both the right-hand and the left-hand regions are very little disturbed by the increase $\Delta n$ of the refractive index.  For normalization purposes we will assume that the  right-hand region has a definite length $L$. It means that the wave number $k$ becomes discrete,
   \begin{equation}
   k_m = \frac{2\pi m}{L}, \quad m \in \langle -\infty, \infty \rangle.
   \end{equation}
   Consider first $t<0$. Following  the same method as in Refs.~\cite{brevik00} and \cite{brevik95}, we will regard  the field component $E_x(x) \equiv E_x(z,t)$ as a scalar quantum field. It is then convenient to change the notation, so  we will write it as $\phi(x)$ instead of as $E_x(x)$. We make the expansion
   \begin{equation}
   \phi(x)= \sum_{m} [a_m u_m(x)+ a_m^\dagger u_m^*(x)], \label{expansion}
   \end{equation}
where
 $a_m$ and $a_m^\dagger$ satisfy the commutation rules
 \begin{equation}
 [a_m, a_{m'}^\dagger]= \delta_{mm'}.
 \end{equation}
The mode functions $u_m(x)$ are  proportional to $\exp[i\Phi_m]$, where
\begin{equation}
\Phi_m= k_m z-\omega_mt,  \quad \omega_m = \frac{n^2\Delta \beta}{\kappa c}k_m. \label{phase}
\end{equation}
The mode functions $u_m$ will be normalized starting from  the Klein-Gordon scalar product,
\begin{equation}
(\phi_1,\phi_2)= -i\int_0^L [\phi_1\partial_t \phi_2^* -(\partial_t\phi_1)\phi_2^*]dz. \label{kleingordon}
\end{equation}
With
\begin{equation}
u_m= \frac{1}{\sqrt{2|\omega_m|L}} \exp[i\Phi_m], \label{mode}
\end{equation}
we then obtain orthonormality for the mode functions,
\begin{equation}
(u_m, u_{m'}) = \delta_{mm'}, \quad  (u_m, u_{m'}^*) = 0.
\end{equation}
Consider next $t>0$, focusing  on the right-moving modes in the right-hand region. There occurs a significant change in the scalar field at $t=0$.  The discrete wave numbers are $k_n=2\pi k_m/L$ as above.
We now write the expansion as
\begin{equation}
\phi(x) =  \sum_{m} [a_{\nu m} u_{\nu m}(x)+ a_{\nu m}^\dagger u_{\nu m}^*(x)],
\end{equation}
where we have introduced an extra subscript $\nu$ on   $a_{\nu m}$ and $u_{\nu m}$ to make clear that they refer to the $t>0$ case. The commutation rules are, similarly as above,
\begin{equation}
 [a_{\nu m}, a_{\nu m'}^\dagger]= \delta_{mm'}.
 \end{equation}
 and the mode functions are proportional to $\exp[i\Phi_{\nu m}]$, where now the phase becomes
\begin{equation}
\Phi_{\nu m}= k_m z-\omega_{\nu m}t,  \quad \omega_{\nu m} =  -\frac{\Delta n}{\kappa c}k_m. \label{nyfase}
 \end{equation}
Thus $\Phi_{\nu m}$ is significantly different from $\Phi_m$ given by Eq.~(\ref{phase}) above. The change in sign of the eigenfrequencies implies that we  will change the sign of the conventional Klein-Gordon product (\ref{kleingordon}) when $t>0$. To summarize, we normalize the mode functions according to
\begin{equation}
(u_m, u_{m'})= -i\int_0^L[ u_m\partial_tu_{m'}^* -(\partial_tu_m)u_{m'}^*] dz, \quad t<0,
\end{equation}
\begin{equation}
(u_{\nu m}, u_{\nu m'})=
 +i\int_0^L[ u_{\nu m}\partial_t u_{\nu m'}^* -(\partial_t u_{\nu m})u_{\nu m'}^*] dz, \quad t>0.
\end{equation}
The new mode functions
\begin{equation}
u_{\nu m}= \frac{1}{\sqrt{2|\omega_{\nu m}|L}}\exp[i\Phi_{\nu m}] \label{nymode}
\end{equation}
then satisfy
\begin{equation}
(u_{\nu m}, u_{\nu m'}) = \delta_{mm'}, \quad  (u_{\nu m}, u_{\nu m'}^*) = 0.
\end{equation}
The scalar fields will be required to be continuous at $t=0$.
We now impose a Bogoliubov transformation relating the $t>0$ mode functions
 to the $t<0$ ones,
\begin{equation}
u_{\nu m}(x)= \sum_{m'} [\gamma(\nu m|m')u_{m'} (x)+\delta(\nu m|m')u_{m'}^* (x)], \label{bogoliubov}
\end{equation}
where $\gamma$ and $\delta$ are coefficients. For the operators we have analogously
\begin{equation}
a_{\nu m}= \sum_{m'} [\gamma(\nu m|m')a_ { m'}+\delta^*(\nu m|m')a_{ m'}^\dagger ].
\end{equation}
From Eq.~(\ref{bogoliubov}) we get
\begin{equation}
\delta(\nu m|m')=-(u_{\nu m}, u_{m'}^*),
\end{equation}
and the number of produced particles in mode $m$ is
\begin{equation}
N_m= \sum_{m'} |\delta(\nu m|m')|^2.
\end{equation}
We need only the modulus of the coefficient $\delta$,
\begin{equation}
|\delta(\nu m|m')|= \big|\int_0^L [u_{\nu m}\partial_t u_{m'} -(\partial_tu_{\nu m})u_{m'}]dz\big|. \label{delta}
\end{equation}
As we consider particle production around the instant $t=0$, we need to consider only the times $t=0-$ and $t=0+$ on each side of the barrier. To make the analysis physically meaningful, we require that the mode functions are continuous at the barrier. As shown from Eqs.
~(\ref{mode}) and (\ref{nymode}), this means that the magnitude of the frequencies coincide at this instant, $|\omega_m| = |\omega_{\nu m}|$ at $t=0$. In turn this means that we are considering only the case where the velocity defect $\Delta \beta$ satisfies the condition
\begin{equation}
 \Delta \beta = \frac{\Delta n}{n^2}. \label{condition}
 \end{equation}
From Eq.~(\ref{delta}),
\begin{equation}
|\delta(\nu m|m')|= \frac{1}{2}\frac{1}{\sqrt{|\omega_{\nu m}\omega_{m'}|}}\, |\omega_{m'}-\omega_{\nu m}|\delta_{m', -m}.
\end{equation}
In this expression  the frequencies $\omega_{m'}$ and $\omega_{\nu m}$ are  in general  not equal at $t=0$. However,   the delta function  requires that $m'= -m$. From Eq.~(\ref{phase}) we  observe that
\begin{equation}
 \omega_{-m}= -\frac{n^2\Delta \beta}{\kappa c}k_m,
 \end{equation}
because $k_m$ is proportional to $m$. Now compare this  equation  with Eq.~(\ref{nyfase}). As
 $n^2\Delta \beta = \Delta n$, it is seen that $\omega_{-m}=\omega_{\nu m}$.  Thus the coefficient $\delta=0$. The artificial  creation of a black hole horizon by means of the Kerr effect in the case when the deviated  subluminal fluid velocity takes the particular value (\ref{condition}), is not accompanied by any particle production. From a physical viewpoint this is not so surprising after all, since the Poynting vector, as we have seen, is continuous across the barrier. This signifies that the  conditions are milder than those encountered in the sudden creation of a cosmic string, for instance.

 A final comment ought to be made on the fact that we have restricted the above analysis to the right-moving modes in the right-hand region ($z>0$) only. The reason is that it is only these modes that are sensitive to the sudden change in the fluid at $t=0$. The
 changes of these modes are large as we have seen: for $t<0$ the energy flux is directed in the $+z$ direction, while for $t>0$ the energy flux  direction is reversed.  The other modes are only slightly affected by this change in the fluid. Thus the right-moving modes in the left-hand region  ($z<0$) are propagating energy in the $+z$ direction also for $t>0$ (recall that the reflection coefficient at $z=0$ is zero). The  left-moving modes in both regions are also less  affected  (taking them into account might lead to some effect).  The situation is in this way simpler than that usually encountered for instance in black hole quantum physics, where one has to apply the Bogoliubov transformation to all (global) modes (a very readable review of that kind of theory can be found in  Ref.~\cite{brout95}).

\end{document}